\documentclass[a4paper,11pt]{article}
\usepackage{jcappub} 
\usepackage{lineno}
\usepackage{bm}
\usepackage{amsmath}
\usepackage[dvipsnames]{xcolor}
\usepackage[normalem]{ulem}
\DeclareUnicodeCharacter{2212}{\ensuremath{-}}
\DeclareUnicodeCharacter{3B5}{$\epsilon$}

\newcommand{\et}{{\bm{\eta}}}
\newcommand{\thet}{{\bm{\theta}}}
\newcommand{\varthet}{{\bm{\vartheta}}}
\newcommand{\xx}{{\bm{x}}}
\newcommand{\bs}{{\bm{S}}}
\newcommand{\ie}{{\textsl{i.e.}}~}

\renewcommand{\arraystretch}{1.1}
\newcommand{\Euclid}{\textsl{Euclid}}
\newcommand{\Planck}{\textsl{Planck}}
\newcommand{\Swyft}{\texttt{Swyft}}
\newcommand{\lcdm}{$\Lambda$CDM}

\title{Fast likelihood-free inference in the LSS Stage IV era}

\author[a]{Guillermo Franco-Abell\'{a}n,}
\author[b, c]{Guadalupe Cañas-Herrera,} 
\author[d, e]{Matteo Martinelli,} 
\author[a]{Oleg Savchenko,}
\author[d, e]{Davide Sciotti,} 
\author[a]{and Christoph Weniger}

\affiliation[a]{ GRAPPA Institute, Institute for Theoretical Physics Amsterdam, \\University of Amsterdam, Science Park 904, 1098 XH Amsterdam, The Netherlands}

\affiliation[b]{European Space Agency/ESTEC\\
Keplerlaan 1, 2201 AZ Noordwijk, The Netherlands}
\affiliation[c]{Lorentz Institute for Theoretical Physics, Leiden University \\ PO Box 9506, Leiden 2300 RA, The Netherlands}
\affiliation[d]{INAF - Osservatorio Astronomico di Roma, via Frascati 33, 00040 Monteporzio Catone (Roma), Italy}
\affiliation[e]{INFN - Sezione di Roma, Piazzale Aldo Moro, 2 - c/o Dipartimento di Fisica, Edificio G. Marconi, I-00185 Roma, Italy}
\emailAdd{g.francoabellan@uva.nl}
\emailAdd{guadalupe.canasherrera@esa.int}
\emailAdd{matteo.martinelli@inaf.it}
\emailAdd{o.savchenko@uva.nl}
\emailAdd{davide.sciotti@inaf.it}
\emailAdd{c.weniger@uva.nl}

\abstract{
Forthcoming large-scale structure (LSS) Stage IV surveys will provide us with unprecedented data to probe the nature of dark matter and dark energy.
However, analysing these data with conventional Markov Chain Monte Carlo (MCMC) methods will be challenging, due to the increase in the number of nuisance parameters and the presence of intractable likelihoods. In light of this, we present the first application of Marginal Neural Ratio Estimation (MNRE) (a recent approach in simulation-based inference) to LSS photometric probes: weak lensing, galaxy clustering and the cross-correlation power spectra. In order to analyse the hundreds of spectra simultaneously, we find that a pre-compression of data using principal component analysis, as well as parameter-specific data summaries lead to highly accurate results. Using expected Stage IV experimental noise, we are able to recover the posterior distribution for the cosmological parameters with a speedup factor of $\sim 10-60$ compared to classical MCMC methods. To illustrate that the performance of MNRE is not impeded when posteriors are significantly non-Gaussian, we test a scenario of two-body decaying dark matter, finding that Stage IV surveys can improve current bounds on the model by up to one order of magnitude. This result supports that MNRE is a powerful framework to constrain the standard cosmological model and its extensions with next-generation LSS surveys.\\ 

\noindent\texttt{GitHub:} We use an implementation of the MNRE algorithm via the open-source code \texttt{Swyft}, available \href{https://github.com/undark-lab/swyft}{here}. Demonstration on Stage IV simulator used in this paper can be found \href{https://github.com/GuillermoFrancoAbellan/Swyft-LSS}{here}. 
}

\begin{document}
\maketitle
\flushbottom

\section{Introduction}\label{sec:intro}

Over the last decades cosmology has transitioned to a precision science: a concordance cosmological model has been established, and its parameters have been measured with exquisite precision \cite{Planck:2018vyg}. This standard cosmological model - also known as \lcdm~- assumes that the universe is composed of around 5\% ordinary matter, 25\% cold dark matter (CDM) and 70\% dark energy in the form of a cosmological constant $\Lambda$. However, the true nature of its main constituents, dark matter and dark energy, still remains elusive. In addition, several observational discrepancies or unresolved questions are starting to accumulate, like the Hubble tension, $S_8$ tension, small-scale CDM crisis, etc (see e.g. \cite{Abdalla:2022yfr} for a recent review). For these reasons, in recent years there has been a growing interest in exploring different extensions of the \lcdm~model, which could shed some light on the mysterious dark components, and possibly offer an explanation for the aforementioned discrepancies.\

Forthcoming LSS Stage IV surveys such as ESA’s \Euclid~satellite mission \cite{EuclidRedBook, Euclid:2019clj} and the Vera C. Rubin Observatory’s Legacy Survey of Space and Time (VRO/LSST) \cite{2009arXiv0912.0201L, LSSTDarkEnergyScience:2018jkl}, are aiming to probe the nature of dark matter and dark energy with unprecedented precision. Nevertheless, the vast data volumes to be delivered by these surveys, as well as the ever-expanding space of theories, motivate a re-thinking of the statistical tools used for parameter inference. Part of the difficulty stems from the fact that classical Bayesian inference methods, such as Markov chain Monte Carlo (MCMC), rely on evaluating the likelihood of the data given the model parameters. Often the likelihood function is only tractable at the level of low-order summary statistics, such as the angular power spectra of weak lensing (WL) and galaxy clustering (GC). Even at the power spectra level, theoretical predictions typically require the output of a Boltzmann solver (like \texttt{CLASS} \cite{Lesgourgues:2011re, Blas:2011rf} or \texttt{CAMB} \cite{Lewis:1999bs, Howlett:2012mh}) together with some recipe to model non-linear scales, which is computationally expensive. Another problem of likelihood-based methods is that they require sampling the \textit{full joint posterior}, which means that the time needed to converge increases dramatically with the dimensionality of the parameter space. This is especially troublesome for extended cosmological models, or in situations where the number of nuisance parameters is very large. For instance, for current galaxy surveys such as the Dark Energy Survey (DES) \cite{DES:2020aks}, the number of free parameters in a typical analysis is of order $\sim30$ \cite{DES:2021wwk}, and this is expected to increase for upcoming surveys such as \Euclid\ and VRO/LSST reaching $\sim50-100$, which means that a typical MCMC can take up to several weeks.

A first attempt to overcome these obstacles is to use trained \textit{emulators}, which allow for ultra-fast likelihood evaluations by efficiently interpolating between a small number of cosmological simulations \cite{Rogers:2018smb, Euclid:2020rfv, SpurioMancini:2021ppk, Gunther:2022pto, Bonici:2022xlo, Nygaard:2022wri}. Since emulating directly the observables is highly inefficient (as these quantities are usually survey dependent), most emulators target the matter power spectrum, which depends on the cosmological model but not on the modelling of systematic effects. Inference can be further optimized if emulators are combined with scalable MCMC methods like Hamiltonian Monte Carlo (HMC), which achieve a highly efficient exploration of the parameter space by exploiting the information coming from the gradient of the likelihood with respect to the parameters \cite{Piras:2024dml}. However, this strategy still comes with some of the caveats inherent to any likelihood-based method, like the need to assume a potentially incorrect functional form of the likelihood (which has to be differentiable for HMC).

Alternatively, novel approaches in simulation-based inference (SBI) allow to perform fast parameter inference without explicitly calculating the likelihood function (see \cite{Cranmer:2019eaq} for a recent review). Instead of a likelihood, the key input in SBI is a \textit{stochastic simulator} that maps from model parameters to realizations of the data\footnote{One can further improve the performance of SBI analyses by accelerating the generation of simulations with the help of emulators. In \autoref{sec:results_ddm} we follow this hybrid approach in order to test a scenario of decaying dark matter.}; this can be seen as sampling an (implicit) likelihood. Furthermore, for many SBI methods the posterior estimates are \textit{amortized}, which allows to perform rigorous statistical consistency checks not accessible to likelihood-based methods, like coverage tests. The field of SBI has recently seen an impressive rate of progress thanks to deep learning methods, and has found application in different astrophysical contexts \cite{Villaescusa-Navarro:2021pkb, Villaescusa-Navarro:2021cni, Zhao:2021ddh, Dax:2021tsq, Crisostomi:2023tle}. 
There are multiple methods for SBI, such as Approximate Bayesian Computation (ABC) \cite{RUBIN_abc, Toni2009ApproximateBC}, Neural Likelihood Estimation (NLE) \cite{papamakarios_nle}, Neural Posterior Estimation (NPE) \cite{papamakarios_npe, Jeffrey:2020itg} and Neural Ratio Estimation (NRE) \cite{Cranmer:2015bka, nre_louppe}. While both NPE and NLE require an estimation of the normalized probability density \cite{Alsing:2018eau, Alsing:2019xrx}, NRE turns inference into a simple binary classification task, thus allowing much more flexibility in its network architecture.

In this work, we consider a variant of NRE called Marginal Neural Ratio Estimation (MNRE), implemented via the open-source code \Swyft\footnote{\href{https://github.com/undark-lab/swyft}{\texttt{https://github.com/undark-lab/swyft}}.}~\cite{Miller:2021hys}. MNRE uses the output of a simulator to train neural networks that directly learn \textsl{marginal} posterior-to-prior ratios. Since MNRE directly achieves 1- and 2-dimensional marginal posteriors, without needing to sample the full joint posterior, it can be significantly more efficient than traditional likelihood-based methods (as well as other SBI techniques) \cite{Cole:2021gwr}. Additionally, MNRE offers the flexibility to ignore large numbers of nuisance parameters, targeting only
the parameters of interest. MNRE has already been applied to a large variety of problems in astrophysics and cosmology, like analyses of the CMB \cite{Cole:2021gwr} and 21-cm signal \cite{Saxena:2023tue}, type Ia supernovae \cite{Karchev:2022xyn, Karchev:2024stw}, strong lensing images \cite{Montel:2022fhv, Coogan:2022cky}, point source searches \cite{AnauMontel:2022ppb},  stellar streams \cite{Alvey:2023pkx} or gravitational waves \cite{Bhardwaj:2023xph, Alvey:2023naa, Alvey:2023npw}.  \

We show the application of MNRE to accelerate parameter inference from upcoming Stage IV photometric galaxy surveys, such as \Euclid~or LSST. While other works have applied different SBI methods to optimize the inference from weak lensing probes \cite{Taylor:2019mgj, Jeffrey:2020xve, Fluri:2022rvb, Lin:2022ayr, DES:2024xij} as well as galaxy clustering probes \cite{Lemos:2023myd, Hahn:2023kky, Hou:2024blc, Modi:2023drt, Tucci:2023bag, Nguyen:2024yth}, our work is the first to use SBI to jointly analyze the hundreds of cosmic shear and galaxy clustering spectra that will be measured by Stage IV surveys. Furthermore, we show that MNRE enables a significant reduction in computational time by doing an explicit comparison with two different MCMC methods (Metropolis-Hastings and Nested Sampling), and derive novel forecasts on a non-standard scenario of two-body decaying dark matter.

The paper is structured as follows. In \autoref{sec:sampling_methods} we discuss the different inference approaches we use in this work, and provide a brief summary of the implementation of MNRE using \Swyft. In \autoref{sec:obs_and_data} we describe the LSS observables that can be constructed from photometric surveys, namely weak lensing and photometric galaxy clustering, detailing our calculation of their angular power spectra in \autoref{sec:theoretical_prediction} and our approach to obtain synthetic data in \autoref{sec:data}. Our results are presented in \autoref{sec:results}. We first describe our inference setup for \Swyft, Metropolis-Hastings and Nested Sampling in \autoref{sec:inference_setup}, and present the forecast posteriors for the $\Lambda$CDM model in \autoref{sec:results_lcdm} and for the two-body decaying dark matter model in \autoref{sec:results_ddm}. We conclude in \autoref{sec:conclusion}. 

\section{Inference methodology}\label{sec:sampling_methods}
The main goal of most cosmological analyses is to obtain constraints on the parameters of the model under investigation, through a comparison of the theoretical predictions for some observable, given by the model, with observational data. To achieve this, a common approach is to reconstruct the full probability distribution of the parameters $\thet$ of a chosen cosmological model $\mathcal{M}$ given the observed data $\xx$, which we denote as $p(\thet|\xx)$. Using the Bayes' theorem, this probability can be expressed as
\begin{equation}\label{eq:BayesTh}
p(\thet|\xx)=\frac{p(\xx|\thet) p(\thet)}{p(\xx)}.
\end{equation}
where $p(\xx|\thet)$ is called \emph{likelihood}, $p(\thet)$ is the \emph{prior}, $p(\xx)$ is the \emph{evidence}, and $p(\thet|\xx)$ is known as \emph{posterior}. The main goal of such an approach is therefore to sample the posterior distribution $p(\thet|\xx)$, as having this probability allows to find the best estimates for the parameters, their marginalised confidence levels and the degeneracies between them. However, there is not a unique approach to sample the posterior probability distribution; one could exploit methods known as \emph{Monte-Carlo Markov Chain} (MCMC), where the parameter space is explored randomly (Monte-Carlo) with each step being dependent only on the previous state (Markov process). This approach can be implemented through different algorithms, such as Metropolis-Hastings \cite{Metropolis:1953am}, Nested Sampling \cite{Skilling:2004pqw}, ensemble sampling \cite{Foreman-Mackey:2012any}, etc. Notice that while the MCMC approach is commonly associated only with the Metropolis-Hastings algorithm, all traditional likelihood-based methods follow the same principles making them part of the MCMC class.\

Nevertheless, we must stress that obtaining the full posterior distribution $p(\thet|\xx)$ is not the real goal of most cosmological analysis. If we are interested in a subset of parameters $\varthet \subset \thet$, what we actually need are the marginalised distribution for each parameter $p(\vartheta_i|\xx)$ in order to obtain confidence level interval on each $\vartheta_i$ separately, or, at most, the marginalised distribution for two parameters $\vartheta_i$ and $\vartheta_j$ in order to visualise their correlations. For such a reason it is beneficial to consider methods returning directly the marginal posteriors.\

In this work we will follow this latter approach, focusing on a recent algorithm in SBI called \emph{Marginal Neural Ratio Estimation} (MNRE), and we will compare its performance with that of MCMC methods. In the rest of this section, we provide more details on the Metropolis-Hastings, Nested Sampling, and MNRE algorithms. We also give a brief summary of the Fisher matrix formalism, since we use this to define our prior boundaries in \autoref{sec:inference_setup}.

\subsection{Fisher matrix}
In order to explore the posterior distribution, most sampling methods require at least an approximate knowledge of the possible values the parameters can take, as well as an initial estimate of the correlation between them. This information can in principle be provided by the results of previous experiments. Here, however, we decide to obtain such information independently, by exploiting the \emph{Fisher matrix} approach \cite{Vogeley:1996xu, Coe:2009xf}. This approach provides a reconstruction of the posterior distribution which relies on assuming that it is Gaussian. The Fisher matrix is defined as the Hessian of the likelihood function and, within the assumption of a Gaussian posterior, coincides with the inverse of the parameters covariance matrix. For two generic parameters $\theta_\alpha$ and $\theta_\beta$, the element of the Fisher matrix $F_{\alpha\beta}$ is defined as
\begin{equation}
F_{\alpha\beta} = -\left. \frac{\partial^2 \mathrm{ln} \mathcal{L}}{\partial \theta_\alpha \partial \theta_\beta} \right\rvert_{\thet^0}\,,
\end{equation}
where $\thet^0$ is the peak of the likelihood distribution $\mathcal{L} = p(\xx|\thet)$, which in the case of forecasts coincides with the assumed fiducial cosmology. 
In the following, we will focus on the LSS observables that can be obtained by photometric surveys, namely weak lensing and galaxy clustering, that we will express in terms of the angular power spectra $C_{ij}^{AB}(\ell)$, where $A$ and $B$ run over the galaxy and lensing fields, while $i$ and $j$ run over the possible combinations of the tomographic redshift bins of the survey. It can be shown that, for such observables, the Fisher matrix can be expressed as \cite{Euclid:2019clj}
\begin{equation}
F_{\alpha\beta} = \sum_{\ell =2}^{\ell_{\mathrm{max}}} \sum_{AB, A'B'} \sum_{ijmn} \frac{\partial C^{AB}_{ij} (\ell)}{\partial \theta_\alpha}  \mathbf{C}^{-1}\left[C_{ij}^{AB}(\ell),C_{mn}^{A'B'}(\ell)\right] \frac{\partial C^{A'B'}_{mn} (\ell)}{\partial \theta_\beta}\,
\label{eq:fisher_matrix}
\end{equation}
where $\mathbf{C}$ is the data covariance matrix. We give more details on how we model this quantity and the observable power spectra in \autoref{sec:obs_and_data}.

\subsection{Likelihood-based inference with Metropolis-Hastings}\label{sec:MH}

The general principle of traditional likelihood-based methods is to generate samples of the full joint posterior distribution $p(\thet|\xx)$, and then marginalize it to get the 1- and 2-dimensional posteriors of interest. With MCMC, this is achieved by starting from a certain point in parameter space $\thet^t$ and selecting the next one $\thet^{(t+1)}$ according to a given proposal distribution $g(\thet^{(t+1)},\thet^t)$. In the Metropolis-Hastings (MH) algorithm, this new sample is ``accepted'' with probability
\begin{equation}\label{eq:MHacceptance}
    p = \mathrm{min} \left\{1,\frac{p(\thet^{(t+1)}|\xx)}{p(\thet^t|\xx)} \right\} .
\end{equation}
We then repeat the cycle, drawing a new proposal based on the last element in the chain. Therefore, the MH algorithm does not only accept new points with higher likelihood, but also points with smaller likelihood at a smaller rate, so that it can explore the parameter space and trace the underlying posterior distribution. In this way, we end generating the weights of each point in parameter space, \ie the number of times we waited and did not move. If the number of steps in the chain is big enough, then these weights are proportional to the posterior $p(\thet|\xx)$. Notice that to exploit the MH algorithm we also need to tune several parameters, such as the starting point of the chains, the prior range for the parameters and the starting proposal distribution, with the choice for these settings significantly affecting the performances of the method \cite{2008ConPh..49...71T}.\ 

In \autoref{sec:results} and \autoref{app:comparison_codes}, we will present cosmological results obtained through the Metropolis-Hastings algorithm as it is implemented in the publicly available Bayesian analysis software for cosmology, \texttt{MontePython}\footnote{\href{https://monte-python.readthedocs.io/en/latest/}{\texttt{https://monte-python.readthedocs.io/en/latest/}}} \cite{Audren:2012wb} and \texttt{Cobaya}\footnote{\href{https://cobaya.readthedocs.io/en/latest/}{\texttt{https://cobaya.readthedocs.io/en/latest/}}} \cite{Torrado:2020dgo}. 

\subsection{Likelihood-based inference with Nested Sampling}

The MH algorithm may face difficulties in efficiently sampling from a multi-modal posterior or one with complicated parameter degeneracies. In addition, the calculation of the Bayesian evidence $p(\xx)$ (a key quantity in model selection) is usually computationally expensive. To address these challenges, alternative likelihood-based approaches such as Nested Sampling have been developed. Nested Sampling algorithms are designed to efficiently compute the evidence, but also produce samples of the posterior $p(\thet|\xx)$ as a by-product. Nested Sampling begins by drawing a set of $n_{\text{live}}$ live points from the prior distribution. At each iteration $i$, the point with the lowest likelihood value $\mathcal{L}_i$ (referred to as a dead point) is replaced with a new point that has a higher likelihood. The prior volume $X(\mathcal{L})$, defined as the fraction of the prior $p(\thet)$ contained within an isocurvature likelihood contour, is given by:
\begin{equation}
    X(\mathcal{L}) = \int_{\mathcal{L}(\thet) > \mathcal{L}} p(\thet)\, d\thet.
\end{equation}
The evidence $p(\xx)$ is then approximated by summing the areas under the $\mathcal{L}(X)$ curve 
\begin{equation}
p(\xx) \approx \sum_{i=1} \mathcal{L}_i w_i,
\end{equation}
where $w_i = (X_{i-1} - X_i)$. Convergence is achieved when the posterior mass $p(\xx)_{\text{live}} \approx \langle \mathcal{L} \rangle_{\text{live}} X_{\text{live}}$ contained by the current set of live points becomes a negligible fraction of the calculated evidence $p(\xx)$. Once the evidence is computed, posterior samples may be obtained by taking the full sequence of dead points and assigning them the weights $p_i = \mathcal{L}_iw_i/p(\xx)$. Different implementations of the Nested Sampling algorithm can be found in publicly available software such as \texttt{PolyChord} \cite{Polychord, Polychord2} and \texttt{MultiNest} \cite{MultiNest}. In this work, we use the latter (integrated in \texttt{MontePython}) for the results presented in \autoref{sec:results}.

\subsection{Simulation-based inference with MNRE}

In simulation-based inference (SBI), the information about the likelihood $p(\xx|\thet)$ is implicitly accessed via a stochastic simulator, mapping from model parameters $\thet$ to data realizations $\xx$. By drawing parameters from the prior $p(\thet)$ and calling the simulator, we can generate $N$ data-parameter pairs $\{(\xx^1,\thet^1),...,(\xx^N,\thet^N)\}$ drawn from the joint distribution $p(\xx,\thet) = p(\xx|\thet) p(\thet)$. Then we can also construct samples from the product of marginal probabilities $p(\xx)p(\thet)$ by randomly shuffling the pair components. The strategy of Neural Ratio Estimation (NRE) is to use these two different sets of data-parameter pairs to train a neural network to approximate the following ratio:
\begin{equation}
r(\xx;\thet) \equiv  \frac{p(\xx,\thet) }{p(\xx)p(\thet)} = \frac{p(\xx|\thet)}{p(\xx)} =\frac{p(\thet|\xx)}{p(\thet)},
\label{eq:ratio_r}
\end{equation}
where the second and third equality follow trivially from Bayes' theorem in \autoref{eq:BayesTh}. In other words, determining $r(\xx;\thet)$ is equivalent to determining the \textit{likelihood-to-evidence} ratio or the \textit{posterior-to-prior} ratio. This will ultimately allow us to obtain posterior samples by drawing samples from the prior $p(\thet)$ and weighting them by $r(\xx;\thet)$. To compute $r(\xx;\thet)$, we optimise a binary classifier $d_\phi (\xx,\thet)$ (with $\phi$ denoting the set of learnable parameters) to distinguish between jointly drawn and marginally drawn pairs, \ie it should be trained such that
\begin{equation}
d_\phi (\xx,\thet) \simeq  
\begin{cases}
1 \ \  \mathrm{if} \ \ (\xx,\thet) \sim p(\xx,\thet),  \\
0  \ \  \mathrm{if} \ \ (\xx,\thet) \sim p(\xx)p(\thet).
\end{cases}
\end{equation}
This learning problem is associated to the minimization of the following loss function
\begin{equation}
\ell[d_\phi(\xx,\thet)] = - \int d\xx d\thet \left[ p(\xx,\thet) \ln  d_\phi (\xx,\thet) + p(\xx)p(\thet) \ln (1-d_\phi (\xx,\thet))  \right],
\label{eq:loss_function}
\end{equation}
also known as binary cross-entropy. Indeed, by analytically minimizing the above functional $\ell [d_\phi(\xx,\theta)]$ one can show that this yields $d^\star_\phi (\xx,\thet) = \sigma (\ln r (\xx,\thet))$, where $\sigma(x) = [1+\mathrm{exp}(-x)]^{-1}$ denotes the sigmoid function. In practice, to find the optimal parameters $\phi$ of the network we minimize a sample-based approximation of the loss function in \autoref{eq:loss_function} using \textit{stochastic gradient descent}. \ 

This method allows to directly estimate marginal posteriors by omitting parameters from network's input, a variant called Marginal Neural Ratio Estimation (MNRE) \footnote{Notice that in MNRE the marginalization is always done implicitly by varying all parameters $\thet$ when generating the simulations.}. More precisely, if $\varthet$ denote the parameters of interest, which are usually a low-dimensional subset of the full parameter set $\varthet \subset \thet = (\varthet,\et)$, we can directly estimate $r(\xx;\varthet) = p(\varthet|\xx)/p(\varthet)$. In practice,  we will only be interested in obtaining 1- and 2-dimensional marginal posteriors, \ie $\varthet = \theta_k$ for some $k$ or $\varthet = (\theta_i, \theta_j)$ for some $(i,j)$. For more details about the algorithm, we refer the reader to \cite{Miller:2021hys} and  \cite{Cole:2021gwr}.

\section{Large-Scale Structure observables and synthetic data}\label{sec:obs_and_data}

The main purpose of this work is to compare the performances of different analysis pipelines in inferring constraints on cosmological parameters from upcoming LSS observations. One of the fundamental ingredients to achieve this goal is the possibility to simulate the observational data we expect to obtain from such surveys. In this section, we provide details on our modelling of a particular set of LSS observables, the assumptions we do to obtain theoretical predictions, and the calculation of the observational noise that we use to obtain synthetic data for LSS surveys.

\subsection{Theoretical Predictions}\label{sec:theoretical_prediction}
For this work, we focus on the main observables that can be obtained from a photometric galaxy survey, namely weak lensing (WL), \ie the correlation between shear in the galaxy shapes, and galaxy clustering (GCph), \ie the correlation between galaxy positions. Together with these, we also consider the cross-correlation of the two observables (XC), and we will refer to the full combination as the 3x2pt statistics. 

We choose to obtain theoretical predictions on these correlations by modelling their angular power spectra in multipole space, and dividing our observations in redshift bins in order to exploit the tomographic information. These angular power spectra can be obtained, applying the Limber approximation \cite{Limber:1954zz}, as
\begin{equation}\label{eq:generic_cl}
    C^{AB}_{ij}(\ell) = \int{\frac{{\rm d}z}{H(z)r^2(z)} W_i^A(z)W_j^B(z)P_{AB}(k_\ell(z),z)}\,
\end{equation}
where the indices $A,B$ run over the two observables (GCph and WL) that we denote respectively with ${\rm g}$ and $\epsilon$, the indices $i,j=1,...,N_{\mathrm{bin},z}$ run over redshift bins, $H(z)$ is the Hubble parameter, $r(z)$ the comoving distance, $W_i^A(z)$ is the kernel function of the considered observable, $P_{AB}(k,z)$ is the power spectrum corresponding to the considered observable combination and $k_\ell(z)=(\ell+1/2)/r(z)$.

The GC term is modelled through the kernel $W_i^{\rm g}(z)$ and the power spectrum $P_{\rm gg}$
\begin{align}
 W_i^{\rm g}(z) &= n_i^{\rm g}(z)H(z)\,,\label{eq:galkernel}\\
 P_{\rm gg}(k,z) &= b_{\rm g}^2(z)P_{\delta\delta}(k,z)\,,\label{eq:galpower}
\end{align}
where $n_i^{\rm g}(z)$ is the normalized distribution in redshift of the observed galaxies in the $i$-th redshift bin, $P_{\delta\delta}(k,z)$ is the matter power spectrum and $b_{\rm g}(z)$ is the galaxy bias, which accounts for the fact that we are observing galaxies, \ie a biased tracer of the total matter distribution. In order to account for uncertainties in galaxy formation and its relation to the underlying distribution of matter, we do not attempt to model the galaxy bias, but rather treat it as a free nuisance parameter $b_i$ assumed to be constant in each redshift bin.

The WL term is instead able to trace directly the total matter distribution, as the observable is the shear of galaxy shapes due to the lensing effect caused by the matter distribution between the observer and the galaxies. However, this observable is also affected by astrophysical systematics. As the shear is obtained by observing galaxy ellipticities, we need to account for effects that can introduce correlations that are not of cosmological origin. Such an effect is usually referred to as Intrinsic Alignment (IA) \cite{Joachimi:2015mma,Kiessling:2015sma,Kirk:2015nma}. We include this effect by defining the WL angular power spectra as \cite{Euclid:2019clj}
\begin{equation}\label{eq:Cl-WL}
    C_{ij}^{\epsilon\epsilon}(\ell)=C_{ij}^{\gamma \gamma}(\ell) + C_{ij}^{{\rm I} \gamma }(\ell) + C_{ij}^{\rm II}(\ell),
\end{equation}

where $\gamma$ refers to the cosmological shear field and $I$ are the contributions due to IA. The terms entering \autoref{eq:Cl-WL} can be obtained through \autoref{eq:generic_cl}, modelling the shear and IA kernels as \cite{Euclid:2019clj}

\begin{align}
    W_{i}^{\gamma}(z) &= \frac{3}{2} H^2_0
\Omega_{{\rm m}} (1 + z) r(z) \int_{z}^{z_{\rm max}}
{{\rm d}z^{\prime} n_{i}^\gamma(z^{\prime}) \left [ \frac{r(z^{\prime}) - 
r(z)}{r(z^{\prime})} \right ]}\,,\label{eq:lenskernel}\\
 W_i^{\rm I}(z) &= n_i^\gamma(z)H(z)\,,\label{eq:IAkernel}
\end{align}
where $H_0$ is the Hubble constant, $\Omega_{\rm m}$ the present-day matter abundance of the Universe, and $n_i^\gamma(z)$ is the normalized distribution in the $i$-th redshift bin for the galaxies used to measure the shear effect. The power spectra entering these terms can be modelled as
\begin{align}
    P_{\gamma\gamma}(k,z) &= P_{\delta\delta}(k,z)\,,\label{eq:lenspower}\\
    P_{\rm II}(k,z) &= f_{\rm IA}^2(z)P_{\delta\delta}(k,z)\,,\label{eq:IApower}
\end{align}
where the first equation shows that shear is an unbiased tracer of the total matter distribution, and the second equation includes the $f_{\rm IA}(z)$ term modelling IA contribution. For this term, we exploit the 
redshift-dependent non-linear alignment (zNLA) model \cite{Joachimi:2015mma,Kiessling:2015sma,Kirk:2015nma}, which allows to define \cite{Euclid:2019clj}
\begin{equation}
    f_{\rm IA}(z)=-{A}_{\rm IA}\mathcal{C}_{\rm IA}\frac{\Omega_{\rm m}}{D(z)} (1 + z)^{\eta^{\rm IA}}\,,\label{eq: IA_function}
\end{equation}
where $A_{\rm IA}$ and $\eta_{\rm IA}$ are the free parameters of the IA model, $\mathcal{C}_{\rm IA}=0.0134$ and $D(z)$ is the growth factor of matter perturbations. With these equations in hand, we are now able to obtain theoretical predictions for our observables as
\begin{align}
C_{ij}^{\epsilon\epsilon}(\ell) =  &
\int{\frac{{\rm d}z}{H(z) r^2(z)}
\left[ W_{i}^{\gamma}(z)W_{j}^{\gamma}(z)
P_{\gamma\gamma}(k_\ell(z),z)+W_{i}^{\rm I}(z)W_{j}^{\gamma}(z)
P_{\rm I\gamma}(k_\ell(z),z)\right.}\nonumber\\
& +\left. W_{i}^{\rm I}(z)W_{j}^{\rm I}(z)
P_{\rm I I}(k_\ell(z),z)\right ]\,,\label{eq:WL_final}\\
C_{ij}^{\rm gg}(\ell) = & \int{\frac{{\rm d}z}{H(z) r^2(z)}W_{i}^{\rm g}(z)W_{j}^{\rm g}(z)
P_{\rm gg}(k_\ell(z),z)}\,,\label{eq:GC_final}\\
C_{ij}^{\rm g\epsilon}(\ell)= & \int \frac{{\rm d}z}{H(z)r^2(z)}\left[ W_{i}^{\rm g}(z) W_{j}^{\gamma}(z)P_{\rm g\gamma}(k_{\ell}(z), z) + W_{i}^{\rm g}(z)W_{j}^{\rm I}(z)P_{\rm gI}(k_{\ell}(z), z)   \right ], \label{eq:XC_final}
\end{align}
with the mixed power spectra being
\begin{align}
    P_{\rm g\gamma}(k,z) &=  b_{\rm g}(z)P_{\delta\delta}(k,z)\,,\\
    P_{\rm gI}(k,z) &=  b_{\rm g}(z)f_{\rm IA}(z)P_{\delta\delta}(k,z)\,,\\
    P_{\rm I\gamma}(k,z) &= f_{\rm IA}(z)P_{\delta\delta}(k,z)\,.
\end{align}

The only ingredients left to be able to compute the $C_{ij}^{AB}(\ell)$ are the cosmological functions entering the expressions, \ie the Hubble parameter $H(z)$, the comoving distance $r(z)$ and the matter power spectrum $P_{\delta\delta}(k,z)$.  Given a set of cosmological parameters, we obtain these from the public Boltzmann solvers \texttt{CLASS} \cite{Lesgourgues:2011re, Blas:2011rf} and \texttt{CAMB} \cite{Lewis:1999bs, Howlett:2012mh}.

\subsection{Synthetic data and fiducial model}\label{sec:data}

\begin{figure}[h!]
    \centering
    \includegraphics[width=0.7\textwidth]{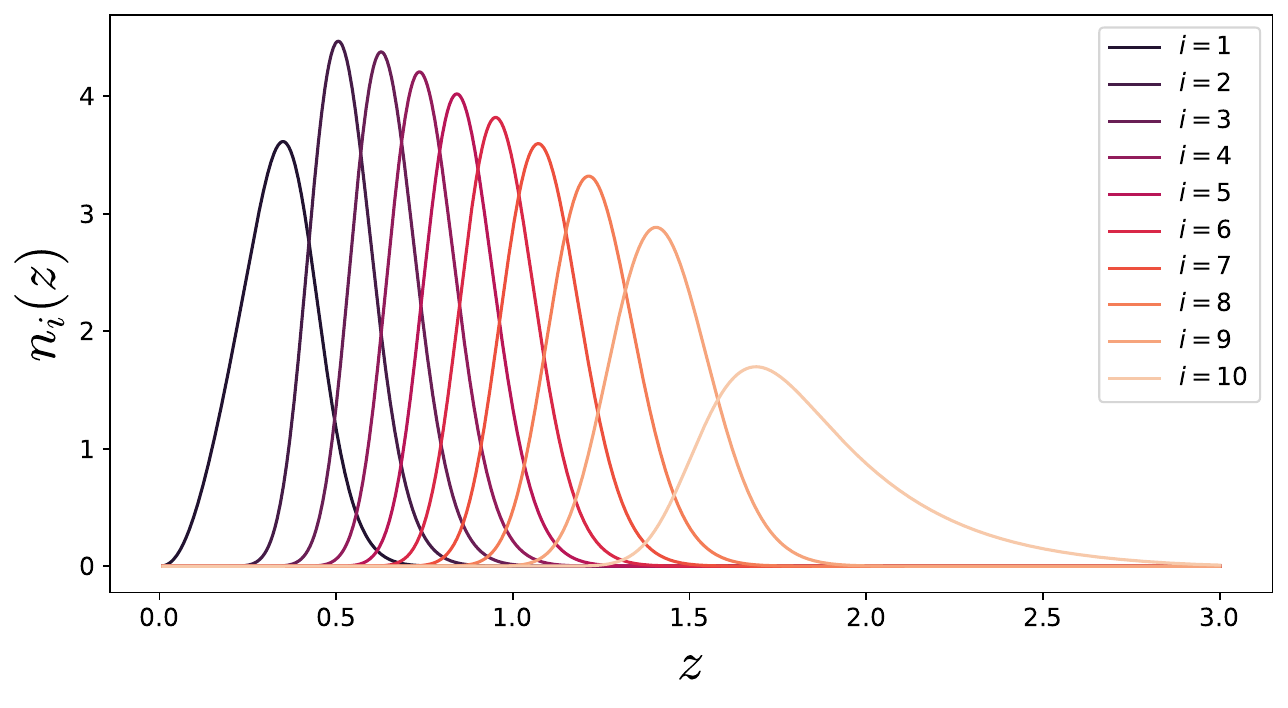}
    \caption{Galaxy redshift distribution $n(z)$ obtained by convolving with the photometric redshift uncertainties using \autoref{eq:binned_dist}. The figure shows $N_{\mathrm{bin},z}=10$ different tomographic redshift bins delimited by redshift values specified in \autoref{eq:binned_redfs}.}
    \label{fig:galdist}
\end{figure}

Once we obtain the theoretical predictions for the observable power spectra, we can use these to simulate synthetic datasets as they would be obtained from an upcoming LSS survey. In order to do so, we can compute predictions for a given fiducial cosmological model that we assume to be the one describing the Universe. The spectra obtained in this way will play the role of observed data, on top of which we can add an observational noise that accounts for the design and specifications of the survey we want to consider.\ 

In this work, we take the flat $\Lambda$CDM model as our fiducial scenario. With this assumption, the parameters we need to choose to obtain the cosmological quantities from the Boltzmann solvers are the baryon and cold dark matter abundances ($\omega_i=\Omega_i h^2$ for $i=\mathrm{b},\mathrm{cdm}$), the Hubble constant ($H_0$) and the amplitude and tilt of the primordial power spectrum of density fluctuations ($A_{\rm s}$ and $n_{\rm s}$) \footnote{Throughout this work, we set two massless and one massive neutrino with fixed mass $m_\nu =0.06 \ \rm{eV}$, following \Planck's conventions \cite{Planck:2018vyg}.}. We report our choice for the fiducial values for these parameters in \autoref{tab:cosmo_settings}.\\

\begin{table}
    \centering
    \begin{tabular}{c|c}
    \hline
        Parameter                      & Fiducial  \\
    \hline
        $H_0$ [km s$^{-1}$ Mpc$^{-1}$] & 67.0 \\
        $100~\omega_{\rm b}$        & 2.2445 \\
        $\omega_{\rm cdm}$            & 0.1206 \\ 
        $n_{\rm s}$                    & 0.96 \\
        $\ln{(10^{10}A_{\rm s})}$      & 3.0569\\
        $A_{\rm IA}$                   & 1.72 \\
        $\eta_{\rm IA}$                & -0.41 \\
        $b_i$                          & $\sqrt{1+\bar{z}_i}$ \\
    \hline
    \end{tabular}
    \caption{Fiducial cosmological and nuisance parameter values used to generate the synthetic dataset.}
    \label{tab:cosmo_settings}
\end{table}

\begin{figure}[h!]
    \centering
    \includegraphics[width=1.0\textwidth]{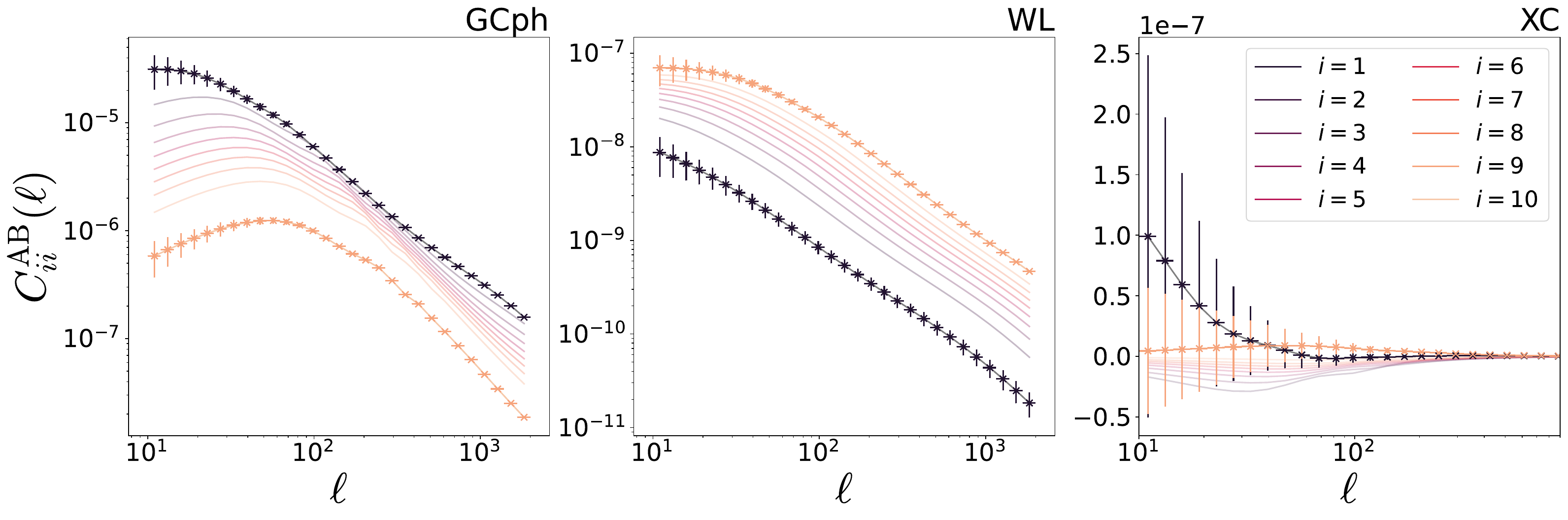}
    \caption{Angular power spectrum $C^{\rm AB}_{ii}(\ell)$ for the photometric LSS probes (3x2pt). From \textit{left} to \textit{right}: photometric galaxy clustering (GCph), weak lensing (WL) and the cross-correlation (XC). For simplicity, we only show the auto-correlation for the 10 tomographic redshift bins corresponding to \autoref{fig:galdist}. For the redshift bins $n_{1}$ and $n_{10}$, we also show the observational errors computed following the discussion in \autoref{sec:noise_and_cov}.}
    \label{fig:cls}
\end{figure}

Furthermore, one key ingredient needed to obtain the $C^{AB}_{ij}(\ell)$ are the binned galaxy distributions $n_i^{\rm g}(z)$ and $n_i^{\gamma}(z)$. For the sake of simplicity, we assume here that the two distributions coincide and we model the unbinned distribution as
\begin{equation}
    n(z)\propto\left(\frac{z}{z_0}\right)^2\exp{\left[-\left(\frac{z}{z_0}\right)^\frac{3}{2}\right]}\,,\label{eq:full_galdist}
\end{equation}
with $z_0=z_m/\sqrt{2}$, where $z_m$, the median redshift of the survey, is reported in \autoref{tab:surveyspecs}. From this, we can obtain the binned distribution, by convolving $n(z)$ with the photometric redshift distribution $p_{\rm ph}(z_{p}|z)$, \ie the probability of measuring a redshift $z_p$ given a true redshift $z$
\begin{equation}
    n_i(z) = \frac{\int_{z_{\rm min}}^{z_{\rm max}}{{\rm d}z_p n(z)p_{\rm ph}(z_p|z)}}{\int_{z_{\rm min}}^{z_{\rm max}}{{\rm d}z\int_{z_-}^{z_+}{{\rm d}z_p n(z)p_{\rm ph}(z_p|z)}}}\,,\label{eq:binned_dist}
\end{equation}
where $z_{\rm min}$ and $z_{\rm max}$ are the limiting redshift of the survey. $z_-$ and $z_+$ are the limiting redshift of the $i$-th bin, and we model $p_{\rm ph}(z_p|z)$ as \cite{Euclid:2019clj}
\begin{align}
    p_{\rm ph}(z_p|z) =& \frac{1-f_{\rm out}}{\sqrt{2\pi}\sigma_b(1+z)}\exp{\left\{-\frac{1}{2}\left[\frac{z-c_bz_p-z_b}{\sigma_b(1+z)}\right]^2\right\}}\nonumber\\
    &+\frac{f_{\rm out}}{\sqrt{2\pi}\sigma_o(1+z)}\exp{\left\{-\frac{1}{2}\left[\frac{z-c_oz_p-z_o}{\sigma_o(1+z)}\right]^2\right\}}\,,
\end{align}
where the set of parameters describing the uncertainties in redshift are shown in \autoref{tab:surveyspecs}. We divide the redshift range of the survey in $N_{\mathrm{bin},z}=10$ equi-populated bins, \ie bins in redshift containing the same number of galaxies, obtained by computing the integral in \autoref{eq:binned_dist}. The set of redshifts delimiting the 10 bins is
\begin{equation}\label{eq:binned_redfs}
    \vec{z}_{\rm lim} = \{0.001,0.417,0.5593,0.677,0.787,0.899,1.017,1.153,1.322,1.573,3.\}\,
\end{equation}
and we show the resulting galaxy redshift distribution $n_i$ in \autoref{fig:galdist}. In addition to the cosmological parameters, we also need to set the fiducial values of the parameters describing the modelling of systematic effects. As indicated in \autoref{tab:cosmo_settings}, for the 10 galaxy bias $b_i$ we take $\sqrt{1+\bar{z}_i}$, where $\bar{z}_i$ are the mean redshifts of each bin, while for the IA parameters we choose values compatible with the literature \cite{Euclid:2019clj}. All these ingredients allow to compute the fiducial angular spectra for the 3x2pt combination of observables, shown in \autoref{fig:cls}.

\subsubsection{Noise and covariance calculation}\label{sec:noise_and_cov}

As we are creating synthetic data that will mimic what we can obtain from observations, a key ingredient to be modelled is the observational uncertainty affecting the measurements, and how this propagates to the angular spectra that we compare with theoretical predictions. For simplicity, we assume that our final measured spectra $C_{ij}^{AB}(\ell)$ are distributed following a multivariate Gaussian distribution, with a covariance matrix $\mathbf{C}\left[C_{ij}^{AB}(\ell),C_{mn}^{A'B'}(\ell')\right]$ that we model as \cite{Euclid:2019clj, 2023arXiv231015731E}
\begin{align}\label{eq:Cls_covariance}
    \mathbf{C}\left[C_{ij}^{AB}(\ell),C_{mn}^{A'B'}(\ell')\right] = & \frac{\delta^{\rm K}_{\ell\ell'}}{(2\ell+1)f_{\rm sky}\Delta\ell}\left\{\left[C_{im}^{AA'}(\ell)+
    N_{im}^{AA'}(\ell)\right]\left[C_{jn}^{BB'}(\ell')+N_{jn}^{BB'}(\ell')\right]+\right.\nonumber\\
    & \left.\left[C_{in}^{AB'}(\ell)+N_{in}^{AB'}(\ell)\right]\left[C_{jm}^{BA'}(\ell')+N_{jm}^{BA'}(\ell')\right]\right\}\,,
\end{align}
where $\delta^{\rm K}_{\ell\ell'}$ is the Kronecker delta function, $f_{\rm sky}$ is the fraction of sky observed by our simulated survey, $\Delta\ell$ the width of the bin in $\ell$ we divide our observations into, and the $A,B$ indexes run over the observables (${\rm g}$, $\epsilon$), while the $(i,j,m,n)$ run over the redshift bins. The information on the observational uncertainties is encoded in the noise terms $N_{ij}^{AB}(\ell)$; we assume this noise to be uncorrelated between different redshift bins and observables and therefore model it as 
\begin{align}
    N_{ij}^{AB}(\ell) = \quad
    \begin{cases}
    \delta^{\rm K}_{ij}/\bar{n}_{\rm g}^i\ \ \ &A=B={\rm g}\,,\\
    \delta^{\rm K}_{ij}\sigma_\epsilon^2/\bar{n}_{\rm g}^i \ \ &A=B=\epsilon\,,\\
    0\ \ \ &A\neq B\,,
    \end{cases}
    \quad    
\end{align}
where $\bar{n}_{\rm g}^i$ is the number of observed galaxies over the survey area in the $i$-th redshift bin and $\sigma_\epsilon$ is the intrinsic ellipticity error, \ie the uncertainty due to the intrinsic shape of the observed galaxies, which is unknown due to our surveys only observing galaxies that have been lensed by the LSS. We show the value used in our synthetic data in \autoref{tab:surveyspecs}. Notice that in the rest of the paper, we will keep the covariance matrix fixed to the value obtained in the fiducial cosmology.

\begin{table}[h!]
    \centering
    \begin{tabular}{|c|c|c|c|c|c|c|c|c|c|c|c|}
         \hline
        $z_m$ & $\bar{n}_{\rm g}$ [arcmin$^{-2}$] & $f_{\rm sky}$ & $\Delta\log_{10}\ell$ & $\sigma_\epsilon$ & $f_{\rm out}$ & $c_b$ & $z_b$ & $\sigma_b$ & $c_o$ & $z_o$ & $\sigma_o$\\ \hline \hline
        0.9 & 30 & 0.35 & 0.08 & 0.3 & 0.1 & 1.0 & 0 & 0.05 & 1.0 & 0.1 & 0.05\\ \hline
    \end{tabular}
    \caption{Parameters describing the survey specifications. All the quantities are defined within \autoref{sec:data}. Notice that in this table we report the value of $\Delta\log_{10}\ell$ rather than $\Delta\ell$, as we choose to perform our analysis using $N_{\mathrm{bin},\ell}=29$ multipole bins, whose limits are logarithmically spaced between $\ell_{\rm min}=10$ and $\ell_{\rm max}=2000$.}\label{tab:surveyspecs}
\end{table}

\begin{figure}[h!]
    \centering
    \includegraphics[width=\columnwidth]{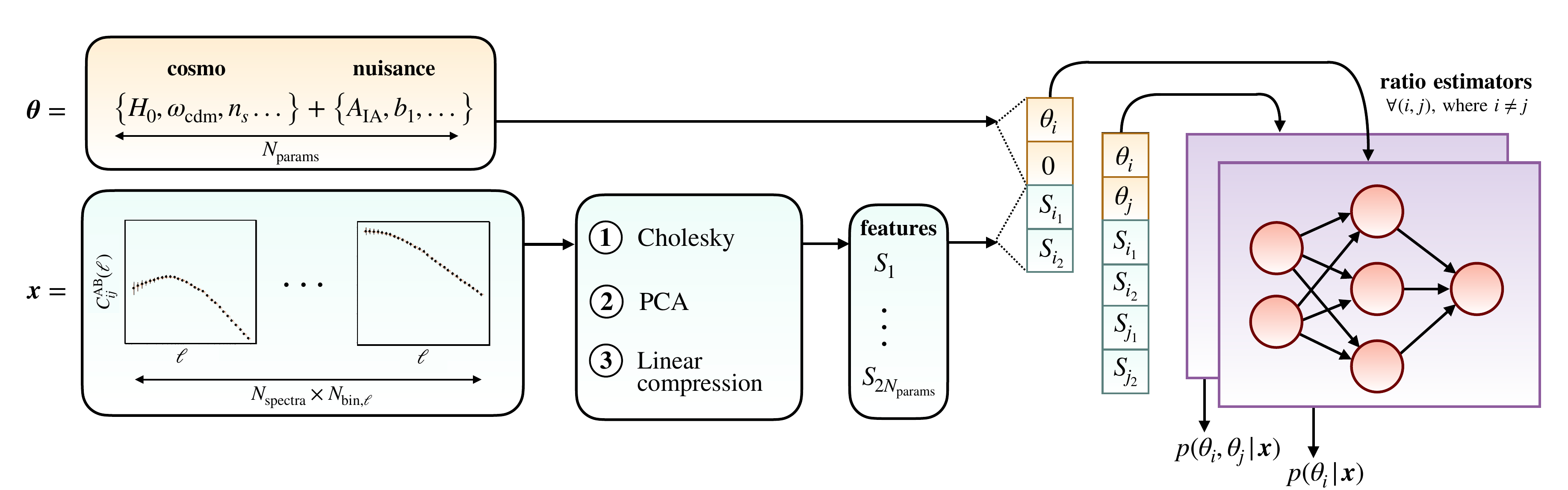}
    \caption{Illustration of the network architecture, mapping from input parameters $\thet$ and data $\xx$ to the 1- and 2-dimensional marginal posteriors of interest. First, the different angular spectra are compressed into features following three steps: i) Cholesky decomposition; ii) Principal Component Analysis; iii) linear compression network. Then, each parameter $\theta_i$ is linked with 2 features $(S_{i_1},S_{i_2})$ and fed as input to the ratio estimators $r(\xx;\theta_i)$ and $r(\xx;\theta_i,\theta_j)$, which are trained with MLPs.   }
    \label{fig:network_diagram}
\end{figure}

\section{Results}\label{sec:results}
\subsection{Inference setup}
\label{sec:inference_setup}
\begin{figure}[h!]
    \centering
    \includegraphics[width=0.8\columnwidth]{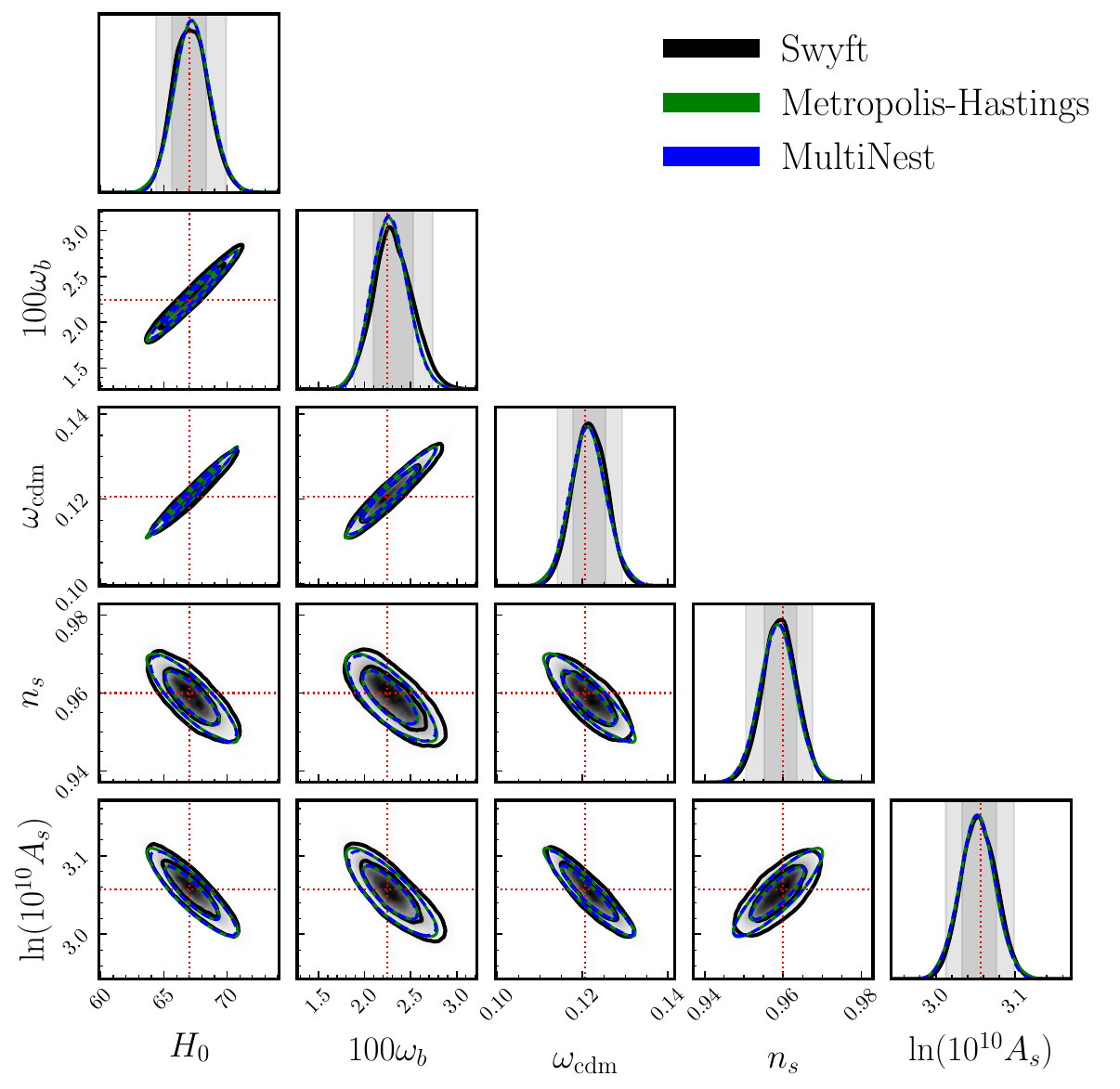}
    \caption{Forecast constraints on the $\Lambda$CDM cosmological parameters from 3x2pt Stage IV photometric probes, using \texttt{Swyft} (black contours), Metropolis-Hastings (green contours) and MultiNest (blue contours). The red dotted lines indicate the injected values. \texttt{Swyft} shows an excellent agreement with classical methods at significantly reduced computational cost.}
    \label{fig:Swyft_vs_mcmc}
\end{figure}
\begin{figure}[h!]
    \centering
    \includegraphics[width=1.0\columnwidth]{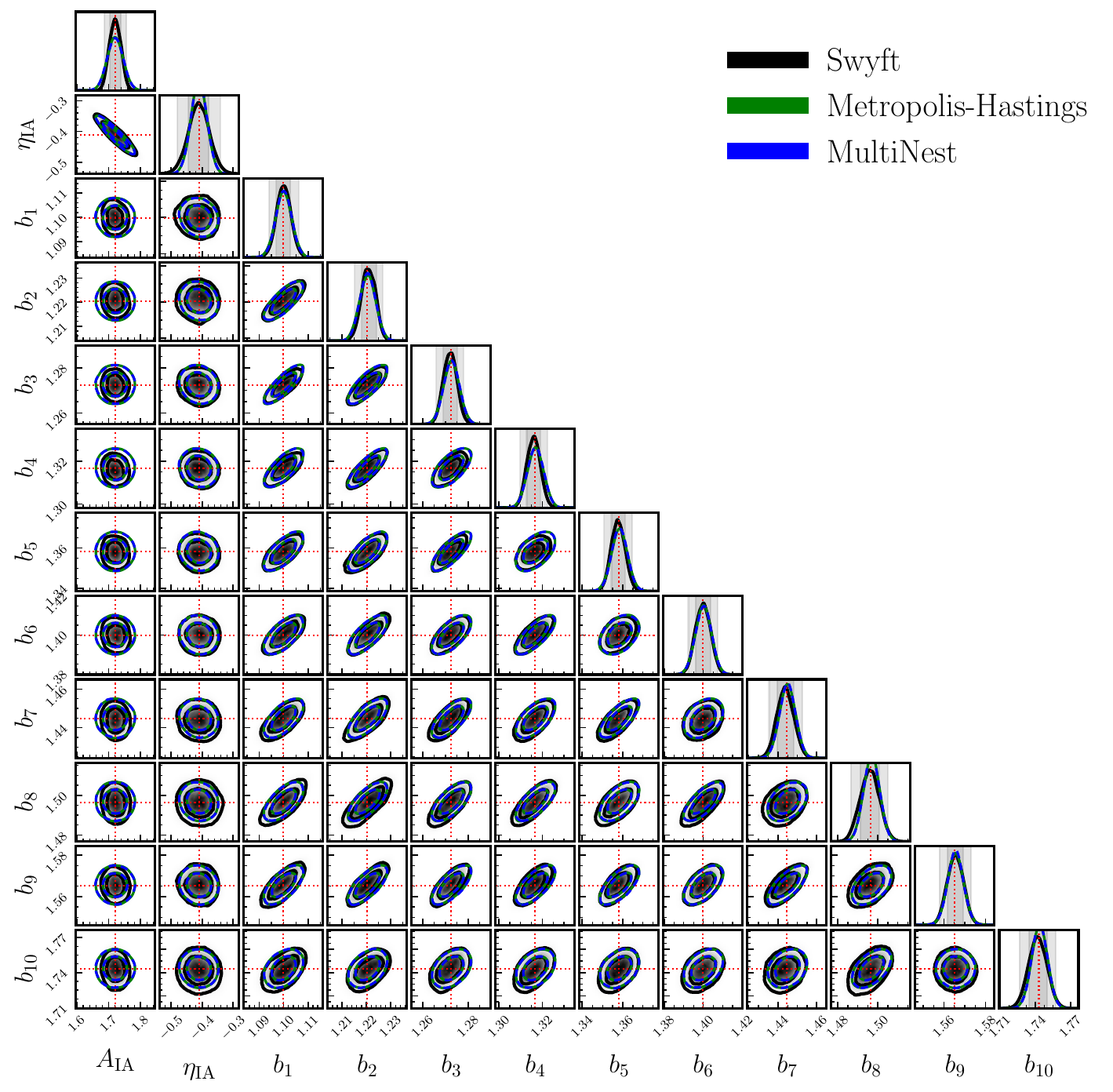 }
    \caption{Forecast constraints on the nuisance parameters from 3x2pt Stage IV photometric probes, using \texttt{Swyft} (black contours), Metropolis Hastings (green contours) and MultiNest (blue contours). The red dotted lines indicate the injected values. }
    \label{fig:Swyft_vs_mcmc_nuisance}
\end{figure}
To perform the \Swyft~analysis we need essentially two ingredients: a forward simulator and a neural network design. A simulator with the same statistical content as the multivariate Gaussian likelihood described in \autoref{sec:noise_and_cov} is defined by the next steps:
\begin{enumerate}
\item Given the cosmological and nuisance parameters, compute each angular spectra $C^{AB}_{ij}$ according to \autoref{eq:WL_final}-\autoref{eq:XC_final}, with the help of a Boltzmann solver. For the analyses shown in this section, we use the public code \texttt{CLASS} \cite{Lesgourgues:2011re,Blas:2011rf} with the \texttt{halofit}\footnote{While there exist more accurate prescriptions to compute the non-linear matter power spectrum, such as \texttt{HMCode} \citep{Mead:2016zqy, Mead:2020vgs}, for this proof-of-concept we have decided to use \texttt{halofit} as this software is included with the same version in both \texttt{CLASS} and \texttt{CAMB}.} prescription \cite{Smith:2002dz} for the non-linear corrections to the matter power spectrum. 
\item Add a noise realization to each angular spectrum, $\hat{C}^{AB}_{ij} = C^{AB}_{ij}+ n^{AB}_{ij}$, where $n^{AB}_{ij}$ is drawn from the multivariate normal distribution $\mathcal{N}(0,\mathbf{C})$, with covariance matrix $\mathbf{C}$ given by \autoref{eq:Cls_covariance}. 
\end{enumerate}
We generate $5\times 10^4$ simulations by varying all cosmological and nuisance parameters,
\begin{equation}
\left\{H_0, 100\omega_b,\omega_{\rm cdm}, n_s,\mathrm{ln}(10^{10}A_s) \right\}+\left\{A_{\rm IA},\eta_{\rm IA}, b_1,...,b_{10} \right\}.
\end{equation}
\noindent
Each parameter $\theta_i$ is drawn from a uniform prior 
\begin{equation}
\theta_i \sim \mathcal{U}([\theta^0_i-5\sigma_i^F,\theta^0_i+5\sigma_i^F]),
\label{eq:prior_range}
\end{equation}
where $\theta^0_i$ denotes the fiducial value in \autoref{tab:cosmo_settings} and the error $\sigma_i^F$ is estimated from the Fisher matrix in \autoref{eq:fisher_matrix} as $\sigma_i^F=\sqrt{(F^{-1})_{ii}}$. We have chosen these reasonably narrow priors just for simplicity. However, we remark that wide priors can easily be accommodated in \Swyft~by using an extension of the algorithm called Truncated Marginal Neural Ratio Estimation (TMNRE) \cite{Cole:2021gwr}, which applies a truncation scheme to quickly zoom into the relevant region of the parameter space.
For example, this could be helpful to study prior volume effects arising in LSS analyses \cite{Simon:2022lde, Holm:2023laa}. We leave this study to future work.\ 

Regarding the design for the network, we split it into a first part performing data compression, and a second part performing the actual inference (see \autoref{fig:network_diagram} for a schematic diagram). More specifically, we found that a pre-compression of data using principal component analysis (PCA), as well as parameter-specific data summaries lead to results with high accuracy and precision. We give more details about the network architecture in \autoref{app:network_architecture}. We use $80\%$ of the simulations as the training dataset, and the remaining  $20\%$ as the validation dataset\footnote{There may be situations where the network tries to ``learn the noise'' in the training set, leading to over-fitting. This is reflected in a decreasing trend in the training loss, but poor performance on the validation loss. To overcome this problem, we found it very convenient to store the spectra and the noise samples separately, and shuffle the noise samples in each batch.}. We use the Adam optimizer with an initial learning rate of $10^{-3}$ and a batch size of $256$. The learning rate is reduced by a factor of $0.1$ whenever the validation loss plateaus for $3$ epochs, and the training is run for no longer than $40$ epochs. The output of the trained network is the different estimated ratios for the parameters of interest.\

On the other hand, we perform the MCMC analyses with the MH and MultiNest algorithms by implementing the aforementioned multivariate Gaussian likelihood in the public code \texttt{MontePython} \cite{Audren:2012wb}. To allow for a fair comparison, we use the same parameters and priors as for the \texttt{Swyft} analysis (see \autoref{eq:prior_range}). For the MH algorithm, we assume chains to be converged using the Gelman-Rubin criterion $R-1<0.01$ \cite{Gelman:1992zz}.  In order to accelerate convergence of MH, for the initial proposal distribution we use the parameter covariance matrix estimated from the inverse of the Fisher matrix, $(F^{-1})_{ij}$. For MultiNest, we take $800$ live points and a tolerance condition on the evidence for stopping the
sampling equal to $0.5$. For both MCMC methods, the chains are post-processed with \texttt{getdist} \cite{Lewis:2019xzd}. \ 

Finally, let us remark that for all analyses carried out in this work we consider a noiseless observation, \ie $\hat{C}^{AB}_{ij} (\thet^0) = C^{AB}_{ij} (\thet^0)$ \footnote{We follow this approach for visualization purposes, so that we obtain posteriors that are centered on the fiducial values. Re-doing our analyses for a noisy observation is perfectly possible (this would give rise to posteriors slightly shifted with respect to the fiducial values), one should just make sure to use the same noise realization in both \Swyft~and MCMC analyses.}. In \autoref{app:comparison_codes} we compare our MH results with those obtained using a different Boltzmann solver (\texttt{CAMB} \cite{Lewis:1999bs, Howlett:2012mh}) and a different sampler (\texttt{Cobaya} \cite{Torrado:2020dgo}), finding good agreement between the different software tools.

\subsection{Forecast posteriors for $\Lambda$CDM}
\label{sec:results_lcdm}

In \autoref{fig:Swyft_vs_mcmc} we show the marginalized 1- and 2-dimensional posteriors of the cosmological parameters from the mock data analysis of 3x2pt Stage IV photometric probes. These have been obtained using MNRE with the \Swyft~code (black contours) and MCMC with the MH and MultiNest algorithms (green and blue contours, respectively). The red dashed lines indicate the true values of the parameters. In \autoref{tab:performance_methods} we additionally show the number of simulator calls (or likelihood evaluations) and total wall-clock time needed for all the analyses carried out in this work. \ 

We find that the \Swyft~and MCMC posterior distributions are in excellent agreement, and both yield a correct reconstruction of the fiducial cosmology. Additionally, using 72 CPU cores, the \Swyft~posteriors were obtained in less than $2$ hours, compared to the $\sim 3$ days (18 hours) required when using MH (MultiNest). This corresponds to a speedup factor of $\sim 10-40$, even after fine-tuning the proposal distribution for MH and using reasonably narrow priors. The substantial reduction in computing time is explained by two factors. First, MNRE directly estimates the marginal posteriors $p(\theta_i|\xx)$ and $p(\theta_i,\theta_j|\xx)$, so it requires far fewer simulation runs than first sampling the full (17-dimensional) joint distribution $p(\thet|\xx)$ and then marginalising. In particular, to achieve converged posteriors for both MCMC methods we needed $\sim 4 \times 10^5$ likelihood evaluations, compared to $5 \times 10^4$ simulations used for \texttt{Swyft}. Secondly, the generation of simulations for MNRE can be made an  embarrassingly parallel operation (and indeed is in our pipeline), whilst this is not possible for typical MCMC methods. We have further checked the statistical consistency of the trained network by evaluating the expected coverage probabilities in \autoref{app:coverage_test}.\ 

Even if the MNRE algorithm gives us the flexibility to ignore large numbers of nuisance parameters, we always have the possibility to quickly estimate their marginal posteriors. With the same simulations that we used to produce \autoref{fig:Swyft_vs_mcmc}, we train every 1- and 2-dimensional marginal posterior of the nuisance parameters used in our analysis. The results are shown in \autoref{fig:Swyft_vs_mcmc_nuisance}. We see that in this case the \Swyft~and MCMC posteriors are in very good agreement, which demonstrates that MNRE can also be used to evaluate consistency and get information about \textsl{e.g.} systematics and astrophysics.

\begin{table}
\centering
\def\arraystretch{1.3}
\begin{tabular}{c|ccc|ccc}
\hline Model & \multicolumn{3}{c|}{\lcdm} & \multicolumn{3}{c}{$\Lambda$DDM }\\ \hline
Method   & \Swyft  & Metropolis   & MultiNest  & \Swyft  & Metropolis  & MultiNest  \\ \hline \hline 
Simulator calls  & $5\times 10^4$   & $4.3\times 10^5$   &  $4.5\times 10^5$ & $3\times 10^4$  & $4.8\times10^5$  & $9.5\times 10^5$ \\
Wall-clock time  & $1.5$h  & $\sim 3$d  & $18$h   & $3$h   & $\sim8$d  & $\sim8$d   \\
\hline
\end{tabular}
\caption{Number of simulator calls and total wall-clock time required for the 3x2pt Stage IV analysis across all the models and methods considered in this work. The wall-clock time for \Swyft~includes both simulation and training, the latter of which only takes $\sim$ 5min on our GPU. The generation of \Swyft~simulations and the Metropolis/MultiNest analyses were performed using the same parallelised hardware (72 CPU cores).}
\label{tab:performance_methods}
\end{table}

\subsection{Forecast posteriors for decaying dark matter}
\label{sec:results_ddm}
\begin{figure}[h!]
    \centering
    \includegraphics[width=0.7\columnwidth]{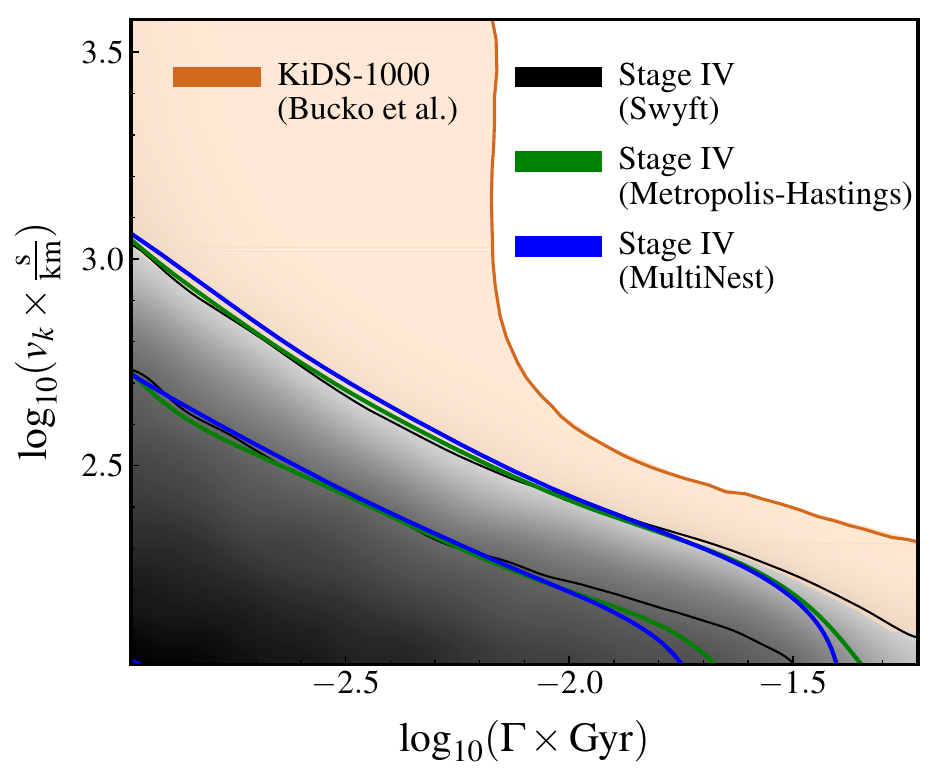}
    \caption{95 $\%$ C.L. bounds on the $\Lambda$DDM parameters $\Gamma$ (decay rate) and $v_k$ (WDM velocity kick). The orange contour denotes the current strongest constraints from KiDS-1000 WL data \cite{Bucko:2023eix}, while the gray contour indicates our projected constraints from 3x2pt Stage IV photometric probes (obtained with \Swyft). For comparison we also show the results obtained with Metropolis-Hastings (green contour) and MultiNest (blue contour). Stage IV surveys can improve the bounds on the $\Lambda$DDM model by up to one order of magnitude.}
    \label{fig:Swyft_vs_mcmc_ddm}
\end{figure}

MNRE can likewise be applied to accelerate the inference of $\Lambda$CDM extensions having significantly non-Gaussian posteriors. To highlight this aspect, we analyze a model (dubbed $\Lambda$DDM) where dark matter is allowed to decay into a massless and a massive particle; the latter of these then behaves in a manner akin to warm dark matter (WDM). The phenomenology of this scenario was reviewed in detail in \cite{FrancoAbellan:2020xnr, FrancoAbellan:2021sxk}, where it was shown to provide a simple way to ease the $S_8$ tension thanks to the time- and scale-dependent power suppression induced by the warm species. Thereafter, the $\Lambda$DDM model was tested using different LSS observables, such as the galaxy power spectrum \cite{Simon:2022ftd}, the Milky Way satellite counts \cite{DES:2022doi}, the Lyman-$\alpha$ forest \cite{Fuss:2022zyt} and the Sunyaev-Zel’dovich galaxy clusters \cite{Tanimura:2023bkh}. Recently, the strongest bounds on this model were derived using the WL power spectra from KiDS-1000 \cite{Bucko:2023eix}. \

This $\Lambda$DDM model is characterized by two extra parameters, the dark matter decay rate ($\Gamma$), and the velocity kick ($v_k$) received by the warm decay product, which for small velocities can be written as
\begin{equation}
\frac{v_k}{c} \simeq \frac{1}{2} \left(1-\frac{m^2_{\mathrm{wdm}}}{m^2_{\mathrm{dcdm}}} \right),    
\end{equation}
where $m_{\mathrm{dcdm}}$ and $m_{\mathrm{wdm}}$ refer to the mass of the dark matter and warm decay product, respectively. Given that $\Lambda$DDM  models with large decay rates and large velocity kicks are already ruled out by observations, we focus on late-time decays ($\Gamma/H_0 \lesssim 1$) and non-relativistic velocity kicks ($v_k/c \ll 1$). In this regime, the expansion history is essentially unaffected, and the main effect is a suppression in the matter power spectrum, whose cut-off scale and amplitude are controlled by $v_k$ and $\Gamma$, respectively \cite{FrancoAbellan:2021sxk}. In order to model the response of dark matter decays on the non-linear power spectrum, \ie $\mathcal{S}(k,z) \equiv P^{\mathrm{nonlin}}_{\Lambda\mathrm{DDM}}(k,z)/P^{\mathrm{nonlin}}_{\Lambda\mathrm{CDM}}(k,z)$, we use the publicly available emulator \texttt{DMemu}\footnote{\href{https://github.com/jbucko/DMemu}{\texttt{https://github.com/jbucko/DMemu}}.}
\footnote{This emulator gives predictions only up to a redshift of $z=2.35$. Since we are exploring slightly larger redshifts in this work, we conservatively set $\mathcal{S}(k,z)=1$ for $z>2.35$. This is anyway expected to have a tiny impact, due to the small weight that redshift distributions give for $z>2$ (see \autoref{fig:galdist}). } that was developed in \cite{Bucko:2023eix}. To perform the \Swyft~and MCMC analyses, we use almost the same setup as that described in \autoref{sec:inference_setup}. This time, we vary not only the $\Lambda$CDM and nuisance parameters, but also the decaying dark matter parameters, using the following priors:
\begin{align}
 \mathrm{log}_{10}(\Gamma \times \mathrm{Gyr}) &\sim \mathcal{U}([-3.0,-1.2]), \\ 
\mathrm{log}_{10}(v_k \times \mathrm{s}/\mathrm{km}) &\sim \mathcal{U}([2.0, 3.6]).
\end{align}
The rest of parameters are varied with somewhat wider priors than in \autoref{eq:prior_range}, namely $\theta_i \sim \mathcal{U}([\theta^0_i-6\sigma_i^F,\theta^0_i+6\sigma_i^F])$, to better capture any potential new degeneracies. The mock observation is generated with the same fiducial cosmology as before, \ie assuming no dark matter decay ($\Gamma, v_k \rightarrow 0$). Using the aforementioned computational resources, we generated $3\times 10^4$ simulations for \Swyft~and trained the marginal posteriors in less than $3$ hours, while the MH (MultiNest) analysis required $5\times 10^5$ ($9 \times 10^5$) likelihood evaluations\footnote{It is interesting to note that the MultiNest analysis required roughly the same number of likelihood evaluations as MH for \lcdm, but almost twice for $\Lambda$DDM (which is adding only two extra parameters). This illustrates the relatively poorer scalability of Nested Sampling with the addition of extra parameters, specially if these introduce complicated degeneracies.} and $\sim 8$ days to fully converge. This corresponds to a speedup factor of $\sim 60$, larger than the one we found for \lcdm. The marginalized 2-dimensional posterior for $\Gamma$ and $v_k$ is shown in \autoref{fig:Swyft_vs_mcmc_ddm}. We can extract two important messages. First, the \Swyft~and MCMC results are again in very good agreement, reinforcing the idea that \Swyft~can easily be applied to test alternative models exhibiting significantly non-gaussian posteriors. Secondly, the comparison with current limits shows that future Stage IV experiments can improve the bounds on the $\Lambda$DDM model by up to 1 order of magnitude. This highlights the great constraining power of surveys like \Euclid~on $\Lambda$CDM extensions predicting a late-time growth suppression, since these surveys will provide very precise tomographic information on the matter power spectrum.

\section{Conclusions and outlook}\label{sec:conclusion}

In this work, we have shown how Marginal Neural Ratio Estimation (MNRE, as implemented in the public code \Swyft) can be used to optimize cosmological parameter inference from upcoming Stage IV photometric galaxy surveys, 
such as \Euclid~or LSST. This technique offers a number of benefits compared to traditional likelihood-based methods:
\begin{enumerate}
\item MNRE enters in the framework of simulation-based inference (SBI) (also known as likelihood-free inference), meaning that it does not need to assume a possibly incorrect functional form for the likelihood, only to generate samples from it. This aspect was not exploited in this work, as the simulated data vectors were sampled from a multivariate Gaussian with given covariance to allow for a direct comparison with MCMC. In the future, we plan to apply MNRE to more realistic forward-simulations of LSS observables capturing any non-Gaussianities in the likelihood, like those recently considered in \cite{vonWietersheim-Kramsta:2024cks}. 

\item  MNRE directly estimates the marginal posteriors of interest, so it can be much more efficient and flexible than conventional MCMC methods (as well as other SBI approaches). This is particularly advantageous when the number of nuisance parameters is very large. Moreover, the generation of simulations can be fully parallelised, and hence made extremely fast given appropriate computational resources.

\item MNRE allows to perform statistical consistency checks which are usually unfeasible for MCMC, such as coverage tests.  
\end{enumerate}
In order to illustrate these aspects, we have focused on the photometric 3x2pt statistics measured at 10 tomographic bins, that we have modelled with a 17-dimensional parameter space (19-dimensional for decaying DM). Our findings can be summarized as follows:
\begin{itemize}
    \item With the training data composed of $5\times 10^4$ angular power spectra and the expected Stage IV experimental noise, we were able to accurately reconstruct the posterior distribution of all $\Lambda$CDM and nuisance parameters with a speedup factor of $\sim 10 (40)$ compared to MultiNest (MH). We also performed coverage tests to validate the behaviour of our estimated posteriors. Our inference pipeline is expected to yield even larger speedup factors for more refined LSS likelihoods/simulators, since an arbitrarily large number of nuisance parameters can be included without increasing the number of required simulations \cite{Cole:2021gwr}. 
    
    \item In order to show that MNRE performs equally well when posteriors are highly non-Gaussian, we have investigated a non-standard model of two-body decaying dark matter that was recently proposed as a solution to the $S_8$ tension \cite{FrancoAbellan:2020xnr}. Using $3\times 10^4$ simulated power spectra as training data, we again find posteriors that are in excellent agreement with MH and MultiNest, this time with a speedup factor of $\sim60$. In addition, we showed that Stage IV surveys will allow to improve current limits on the two free parameters of the model by up to one order of magnitude. This illustrates that MNRE provides a powerful tool to test extended cosmologies, which is often computationally prohibitive with classical methods.
\end{itemize}
It will be interesting to extend our pipeline to include other LSS observables, such as spectroscopic galaxy clustering, and explore how constraints evolve with different combinations of probes. In fact, MNRE gives the opportunity to quickly perform massive global scans, since for a given theoretical model one can re-use simulations to perform inference for many different experimental configurations, while for MCMC one is forced to start new chains every time different observations are considered \cite{Cole:2021gwr}. Finally, let us note that, while in this work we have focused on two-point statistics, these capture only a fraction of relevant cosmological information. In recent years there has been a growing interest in performing inference directly with field-level LSS data, in order to extract all the information available in higher order statistics and better constrain the cosmological parameters \cite{Jasche:2018oym,Andrews:2022nvv,DES:2024xij,Lemos:2023myd,Nguyen:2024yth}. In future work, we plan to use MNRE to perform field-level inference of Stage IV galaxy surveys. 

\section*{Acknowledgements and code availability}

GFA, OS and CW acknowledge support from the European Research Council (ERC) under the European Union's Horizon 2020 research and innovation programme (Grant agreement No. 864035 - Undark).
MM acknowledges funding by the Agenzia Spaziale Italiana (\textsc{asi}) under agreement no. 2018-23-HH.0 and support from INFN/Euclid Sezione di Roma. GCH acknowledges support through the ESA research fellowship programme.  The main analysis for this work was carried out on the Snellius Computing Cluster at SURFsara. The authors would like to thank Stéphane Ilić for his participation in the early stages of this work. The authors also thank James Alvey, Uddipta Bhardwaj and Noemi Anau Montel for useful discussions.\ 

Software used in this work includes \texttt{NumPy} \cite{Harris:2020xlr}, \texttt{Matplotlib} \cite{Hunter:2007ouj}, \texttt{PyTorch} \cite{Paszke:2019xhz}, \texttt{PyTorch Lightning} \cite{falcon_2024_10779019}, \texttt{Joblib} \cite{joblib} and \texttt{Jupyter} \cite{jupyter}. Accompanying code for the examples in this paper is available on \href{https://github.com/GuillermoFrancoAbellan/Swyft-LSS}{\texttt{https://github.com/GuillermoFrancoAbellan/Swyft-LSS}}.

\appendix

\section{Network architecture}
\label{app:network_architecture}

As discussed in \autoref{sec:inference_setup}, we have split the network architecture for the ratio estimation into two separate components: i) a data compression network $\bs = C_\phi (\xx)$, which learns to compress the potentially high-dimensional data $\xx$ into low-dimensional features $\bs$; ii) a network to perform the actual ratio estimation, by learning to discriminate between marginally and jointly drawn model parameters $\thet$ and feature vectors $\bs$. The details of the compression network are dictated by the specific properties of the data. We recall that in our case the data can be written as a vector $\xx = \bf{s}(\thet) + \bf{n}$, where  $\bf{s}(\thet)$ is the concatenation of all angular spectra computed according to \autoref{eq:WL_final}-\autoref{eq:XC_final}, and $\bf{n}$ is a noise vector drawn from the multivariate normal distribution $\mathcal{N}(0,\mathbf{C})$, with  covariance matrix $\mathbf{C}$ given by \autoref{eq:Cls_covariance}. We have a total of $N_{\mathrm{spectra}} = N_{\mathrm{bin},z}(2N_{\mathrm{bin},z}+1) = 210$ distinct power spectra, each binned in $N_{\mathrm{bin},\ell}=29$ multipoles, so the size of the data vector $\xx$ is $6090$. Since this vector contains a large number of correlated noisy spectra, we found beneficial to do the data compression by taking the following steps:
\begin{enumerate}
    \item In order to decouple the different spectra, we start by performing a Cholesky decomposition of the covariance matrix, $\mathbf{C}=\mathbf{L} \mathbf{L}^T$ (where $\mathbf{L}$ is a lower triangular matrix), and change to the basis $\Tilde{\xx} = \mathbf{L}^{-1}\xx$. 
    \item Secondly, we do a Principal Component Analysis (PCA) in order to rotate to a smaller basis capturing the largest data variations. More precisely, we construct a matrix $\mathbf{A}$ by stacking $10^4$ samples of the vector $\Tilde{\bf{s}} = \mathbf{L}^{-1}\bf{s}$, and subsequently perform a singular value decomposition $\mathbf{A} = \mathbf{U}\mathbf{D}\mathbf{V}$, where $\mathbf{U}$ and $\mathbf{V}$ are unitary matrices and $\mathbf{D}$ is a rectangular diagonal matrix. We then project the data to the new basis $\Tilde{\xx}'= \mathbf{V} \Tilde{\xx}$, but retaining only the first $N_{\rm PCA}$ components for which the eigenvalues contribute to more than $0.1\%$ of the total (typically $N_{\rm PCA}\sim 30-40$).
    \item  Thirdly, we use a linear compression network that maps the rotated data vector $\Tilde{\xx}'$ into a feature vector $\bs$. 
\end{enumerate}
Notice that the first two steps are performed only once at the beginning, while for the third step we optimize the parameters of the linear compression network during training. We assume that each parameter can be well described by 2 features, so the feature vector $\bs$ has a size $2N_{\rm params}$, where $N_{\rm params}$ is the number of parameters of interest. We concatenate each of these parameters with the corresponding features, and fed this as input to the ratio estimators, in charge of estimating every 1- and 2-dimensional marginal posterior (see \autoref{fig:network_diagram}). Each ratio estimator is described by a multi-layer perceptron (MLP) with four layers, each containing 64 neurons.

\section{Coverage of the network}
\label{app:coverage_test}
One of the key motivations for using MNRE is to apply it to problems in cosmology where the inference with traditional methods is extremely time-consuming, or even impossible. Hence, it is important to have additional consistency checks that do not require having a ground-truth against which to compare the results (such as the converged MCMC that we showed in \autoref{sec:results}). \ 

\begin{figure}[h!]
    \centering
    \includegraphics[width=\columnwidth]{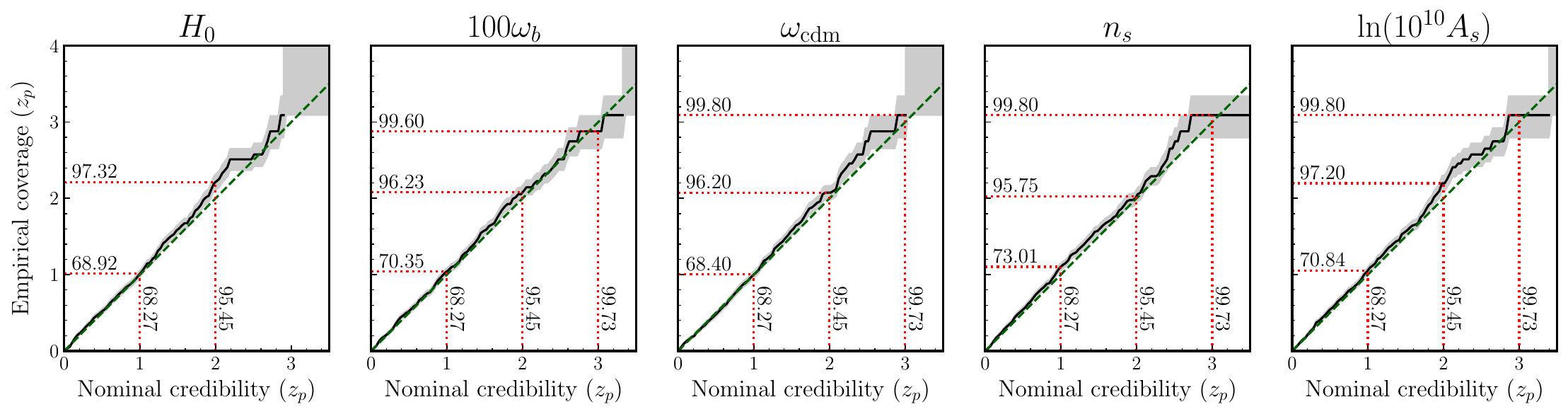}
    \caption{Coverage test for the cosmological parameters from the $\Lambda$CDM  analysis of 3x2pt Stage IV photometric probes. The black solid line indicates the average coverage, while the gray contour indicates the $68\%$ uncertainty on the coverage. We see that the empirical coverage and confidence level generally match to excellent precision.}
    \label{fig:coverage_test}
\end{figure}

In the context of SBI, the most well-established techniques are known as \textit{coverage tests}. These sort of methods exploits the fact that, once the inference networks are trained, they can effortlessly generate the posteriors for \textit{any} mock observation simulated from the prior (as opposed to MCMC, where one gets the posterior for a single observation). Hence, the idea is to perform inference on many different mock data simulations to get the \textit{empirical coverage}, which gives the proportion of time that a certain interval contains the true parameter value. For a well-calibrated posterior, the $p\%$ credible interval should contain the simulation-truth value $p\%$ of the time, so one should get a totally diagonal line when plotting the empirical coverage against the expected credible intervals. These tests are generally unfeasible for sequential MCMC methods such as Metropolis-Hastings, where one needs to rely on convergence criteria like the Gelman-Rubin statistic \cite{Gelman:1992zz}. However, it should be noted that coverage tests provide a necessary \textit{but not sufficient} condition for the calibration of the posteriors. Developing new tests in order to trust results from SBI is an active area of ongoing research \cite{Hermans:2021rqv, Lemos2023}.\ 

Using a batch of 500 simulations, we perform the coverage test for the marginal posteriors of the five $\Lambda$CDM parameters trained in \autoref{sec:results_lcdm}. Instead of showing the highest posterior density region $p$, we choose to work with a new variable $z_p$ defined by $p/100 = \frac{1}{\sqrt{2\pi}}\int_{-z_p}^{z_p}dz \exp{(-z^2/2)}$, to place more emphasis in the tail of the posteriors. This means that the $(1, 2, 3)\sigma$ regions correspond to $z_p = (1, 2, 3)$ with $p = (68.27, 95.45, 99.97)$. We also compute the uncertainty on the empirical coverage arising from the finite sample size, using the Jeffreys interval (see \cite{Cole:2021gwr} for details). The results of the test are shown in \autoref{fig:coverage_test}. We find that for every parameter, we achieve excellent posterior coverage, which adds additional validation to our agreement with MCMC. 

\begin{figure}[h!]
    \centering
    \includegraphics[width=0.7\columnwidth]{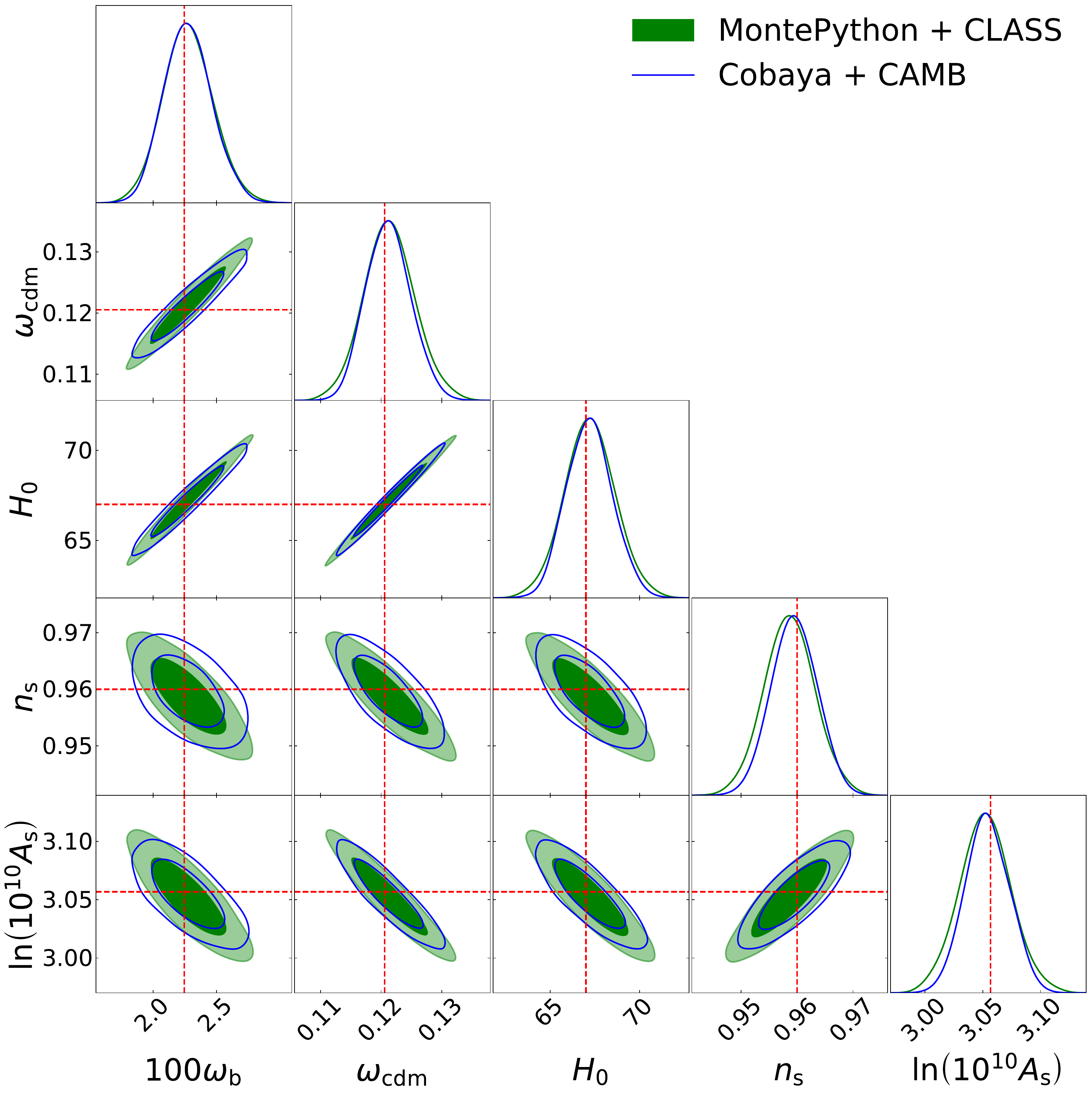}
    \caption{Constraints on the cosmological parameters from 3x2pt Stage IV photometric probes. We show the difference between the two codes using a similar implementation of the Metropolis-Hastings algorithm: \texttt{Cobaya} (blue, with \texttt{CAMB} as the Boltzmann solver) and \texttt{MontePython} (green, with \texttt{CLASS} as the Boltzmann solver). The red dashed lines indicate the injected values.}
    \label{fig:compare_cosmo}
\end{figure}

\begin{figure}[h!]
    \centering
    \includegraphics[width=1.0\columnwidth]{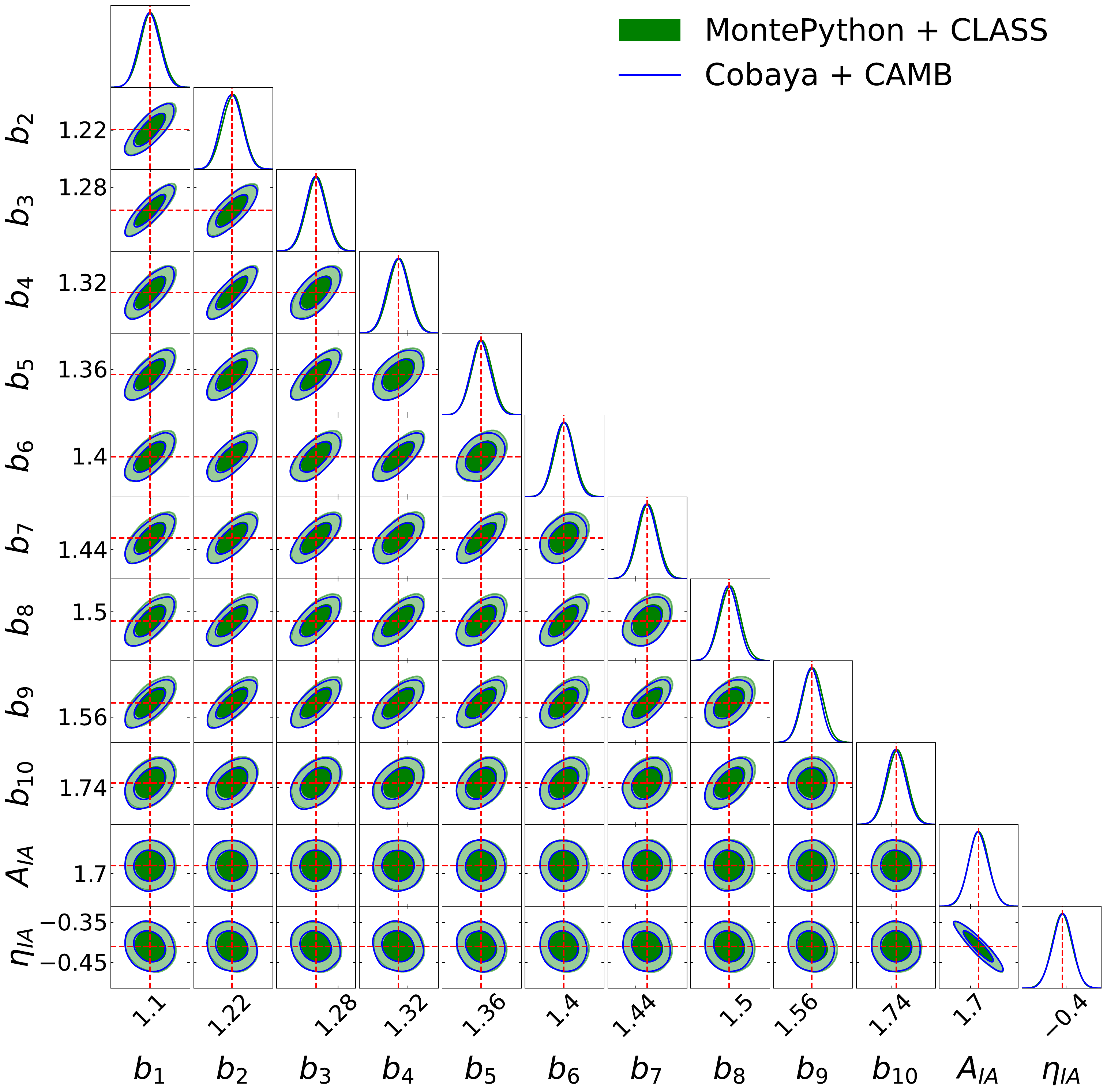}
    \caption{Constraints on the nuisance parameters from 3x2pt Stage IV photometric probes. We show the difference between the two codes using a similar implementation of the Metropolis-Hastings algorithm: \texttt{Cobaya} (blue, with \texttt{CAMB} as the Boltzmann solver) and \texttt{MontePython} (green, with \texttt{CLASS} as the Boltzmann solver). The red dashed lines indicate the injected values. }
    \label{fig:compare_nuis}
\end{figure}

\section{Comparison between \texttt{MontePython+CLASS} and \texttt{Cobaya+CAMB}}
\label{app:comparison_codes}

In order to double-check our methodology, we have conducted the MH analyses using two distinct combinations of software tools: \texttt{MontePython} with \texttt{CLASS} and \texttt{Cobaya} with \texttt{CAMB}. As shown in \autoref{fig:compare_cosmo} and \ref{fig:compare_nuis}, we find excellent agreement between these different codes. This underscores the reliability and robustness of both MCMC pipelines, and reinforces the idea that the main advantages found with \texttt{Swyft} still hold regardless of which MCMC pipeline is used. It also suggests that either combination can be safely utilized for further comparisons with \Swyft~(\ie \texttt{Swyft} can be trained with \textit{simulations} from \texttt{CLASS} or \texttt{CAMB}).

\newpage

\bibliographystyle{JHEP}
\bibliography{biblio}

\end{document}